\documentclass[aps,prb,twocolumn,superscriptaddress,showpacs,floatfix,amsmath,amssymb,10pt]{revtex4-1}

\usepackage{float}
\usepackage{caption}
\usepackage{hyperref}
\usepackage{graphicx}
\usepackage{physics}
\usepackage{titlesec}
\usepackage{float}

\begin{document}

\title{Deep Learning on the 2-Dimensional Ising Model to Extract the Crossover Region with a Variational Autoencoder}

\author{Nicholas Walker}
\affiliation{Department of Physics \& Astronomy, Louisiana State University, Baton Rouge, Louisiana 70803, USA}
\author{Ka-Ming Tam}
\affiliation{Department of Physics \& Astronomy, Louisiana State University, Baton Rouge, Louisiana 70803, USA}
\affiliation{Center for Computation \& Technology, Louisiana State University, Baton Rouge, Louisiana 70803, USA}
\author{M.\ Jarrell}
\affiliation{Department of Physics \& Astronomy, Louisiana State University, Baton Rouge, Louisiana 70803, USA}
\affiliation{Center for Computation \& Technology, Louisiana State University, Baton Rouge, Louisiana 70803, USA}
\affiliation{Deceased July 20th, 2019}

\begin{abstract}
The 2-dimensional Ising model on a square lattice is investigated with a variational autoencoder in the non-vanishing field case for the purpose of extracting the crossover region between the ferromagnetic and paramagnetic phases. The encoded latent variable space is found to provide suitable metrics for tracking the order and disorder in the Ising configurations that extends to the extraction of a crossover region in a way that is consistent with expectations. The extracted results achieve an exceptional prediction for the critical point as well as agreement with previously published results on the configurational magnetizations of the model. The performance of this method provides encouragement for the use of machine learning to extract meaningful structural information from complex physical systems where little \textit{a priori} data is available.
\end{abstract}

\maketitle

\section{Introduction}

Machine learning (ML) and consequently data science as a whole have seen rapid development over the last decade or so, due largely to considerable advances in implementations and hardware that have made computations more accessible. Conceptually, the ML approach can be regarded as a data modeling approach employing algorithms that eschew explicit instructions in favor of strategies based around pattern extraction and inference driven by statistical analysis. This presents a colossal opportunity for modern scientific investigations, particularly numerical studies, as they naturally involve large data sets and complex systems where obvious explicit instructions for analysis can be elusive. Conventional approaches often neglect possible nuance in the structure of the data in favor of rather simple measurements that are often untenable for sufficiently complex problems. Some ML methods such as inference methods have been routinely applied to certain physical problems, such as the maximum likelihood method and the maximum entropy method,\cite{PhysRevB.44.6011,JARRELL1996133} but applications which utilizing ML methods have only recently attracted attention in the physical sciences, particularly for the study of interacting systems on both classical and quantum scales. \cite{carrasquilla2017machine} There is a unique opportunity to take advantage of the advances in ML algorithms and implementations to provide interesting new approaches to understanding physical data and even perhaps improve upon existing numerical methods.\cite{PhysRevB.95.035105} Outstanding problems involving the predictions of transition points and phase diagrams are also of great interest for treatment with ML methods.

In order to utilize ML approaches for studying phase transitions, one must assume that there is some pattern change in the measured data across the phase transition. Fortunately, this is in fact exactly what happens in most phase transitions. The widely adopted Lindemann parameter, for example, is essentially a measure of the deviations of atomic positions in the system from equilibrium positions and is often used to characterize the melting of a crystal structure.\cite{lindemann} Similar form of pattern changes in the positions of the constituent atoms are often present in molecular systems in general. Perhaps more importantly, for some sufficiently complex systems, their phase transitions do not have obvious order parameters, often prohibiting the detection of such pattern changes using conventional methods. This is not a hypothetical situation, indeed hidden orderings for some interesting materials, such as heavy fermion materials and cuprate superconductors, have been proposed for long time. \cite{PhysRevLett.96.036405,cuprate_hidden_order,HF_hidden_order}

Other systems may not even exhibit a true phase transition, but rather a crossover region where there is no singularity across different phases that can be difficult to characterize with conventional methods. A conventional phase transition can be identified in two ways, with the first being a singularity in a derivative of the free energy as proposed by Ehrenfest and the second being a broken symmetry exhibited by an order parameter as proposed by Landau. Unlike a conventional phase transition, a crossover is not identified by a singularity in the free energy. There is also no broken symmetry in such a situation and thus no order parameter is associated with a crossover. The order parameter and singularity in the free energy are presumably sharp and obvious features which can be rather easily identified. The absence of such features clearly present a challenging situation in the prediction of a crossover region by ML. ML is a new route of studying these systems by searching for hidden patterns in the measured data where readily applicable \textit{a priori} information is in short supply.

A viable ML method for detecting a crossover will find its use in many interesting systems related to the quantum phase transition.\cite{Vojta_2003} While the quantum phase transition is a second order phase transition controlled by non-thermal parameters at zero temperature, all experiments and most numerical simulations are conducted at finite albeit low temperatures for practical reasons. As a consequence of said thermal conditions, quantum critical points at low temperatures behave as crossover phenomenona. It is widely believed that many interesting materials, particularly high temperature cuprate superconductors, harbor a quantum critical point. An ML approach for detecting the crossover phenomenon can thus be an important tool for studying quantum critical points. 

Work has been done on various problems to characterize phase transitions in physical systems using ML methods, including the Ising model in the vanishing field case.\cite{carrasquilla2017machine, PhysRevB.94.195105, structure_classification, melting, lattice_spin, ising_vae, ising_vae_trans, ml_phase_trans_cross, ml_order, unsup_trans, ising_crit, ising_boltzmann, ising_boltzmann_2} This work will use a similar approach to those seen in these papers, but will focus on the crossover regions that are introduced in the non-vanishing field case of the 2-dimensional Ising model instead of seeking only the exactly known transition point in the vanishing field case.\cite{PhysRev.65.117} This is a somewhat more difficult problem, as there is no explicit transition to be found, but it remains an interesting problem nonetheless and possibly carries much greater implications for crossover regions in more complicated problems.

The Ising model itself is a mathematical model for ferromagnetism that is often explored in the field of statistical mechanics in physics to describe magnetic phenomena.\cite{p_cond_mat} Originally, the Ising model was developed to investigate magnetic phenomena, as mentioned earlier. With the discovery of electron spins, the model was designed to determine whether or not local interactions between magnetic spins could induce a large fraction of the electronic spins in a material to align in order to produce a macroscopic net magnetic moment. It is expressed in the form of a multidimensional array of spins $s_i$ that represent a discrete arrangement of magnetic dipole moments of atomic spins.\cite{p_cond_mat} The spins are restricted to spin-up or spin-down alignments such that $s_i \in \qty{-1, +1}$. The spins interact with their nearest neighbors with an interaction strength given by $J_{ij}$ for neighbors $s_i$ and $s_j$. The spins can additionally interact with an applied external magnetic field $H_i$ (where the magnetic dipole moment $\mu$ has been absorbed). The full Hamiltonian describing the system is thus expressed as

\begin{equation}
\mathcal{H} = -\sum_{\left< i, j \right>} J_{ij} s_i s_j - \sum_i H_i s_i
\end{equation}

Where $\expval{i, j}$ indicates a sum over adjacent spins. For $J_{ij} > 0$, the interaction between the spins is ferromagnetic, for $J_{ij} < 0$, the interaction between the spins is antiferromagnetic, and for $J_{ij} = 0$, the spins are noninteracting. Furthermore, if $H_i > 0$, the spin at site $i$ tends to prefer spin-up alignment, if $H_i < 0$, the spin at site $i$ tends to prefer spin-down alignment, and if $H_i = 0$, there is no external magnetic field influence on the spin at site $i$. The model has seen extensive use in investigating magnetic phenomena in condensed matter phhysics.\cite{spin_glass_1,spin_glass_2,ising_app_1,ising_app_2,ising_app_3,ising_app_4,ising_app_5} Additionally, the model can be equivalently expressed in the form of the lattice gas model, described by the following Hamiltonian

\begin{equation}
\mathcal{H} = -4J\sum_{\expval{i,j}}n_i n_j - \mu\sum_i n_i
\end{equation}

Where the external field strength $H$ is reinterpreted as the chemical potential $\mu$, $J$ retains its role as the interaction strength, and $n_i \in \Bqty{0, 1}$ represents the lattice site occupancy. The original Ising Hamiltonian can be recovered using the relation $\sigma_i = 2n_i-1$ up to a constant. This model describes a multidimensional array of lattice sites which can be either occupied or unoccupied by a hard shell atom, disallowing occupancy greater than one. The first term is then interpreted as a short-range attractive interaction term while the second is the flow of atoms between the system and the reservoir. This is a simple model of density fluctuation and the liquid-gas transformations used primarily in chemistry, albeit often with modifications.\cite{ising_chem_1,ising_chem_2} Additionally, modified versions of the lattice gas models have been applied to binding behavior in biology.\cite{ising_bio_1,ising_bio_2,ising_bio_3}

Typically, the model is studied in the case of $J_{ij} = J = 1$ and the vanishing field case $H_i = H = 0$ is of particular interest for dimension $d \ge 2$ since a phase transition is exhibited as the critical temperature is crossed. For two dimensions, the critical temperature can be identified by exploiting Kramers-Wannier duality symmetry.\cite{kwd,kwd_2,kwd_3} At low temperatures with a vanishing field, the physics of the Ising model is dominated by the nearest-neighbor interactions, which for a ferromagnetic model means that adjacent spins tend to align with one another. However, as the temperature is increased, the thermal fluctuations will eventually overpower the interactions such that the magnetic ordering is destroyed and the orientations of the spins can be considered independent of one another. This is called a paramagnet.

In such a case, if an external magnetic field were to be applied, the paramagnet would respond to it and tend to align with it, though for high temperatures, a sufficiently strong external field will be required to overcome the thermal fluctuations. Since the magnetization smoothly decreases to zero with increasing temperature in the presence of an external magnetic field, there is no phase transition where the magnetization abruptly vanishes. Instead, the region in which the system goes from an ordered to a disordered state is referred to as the crossover region. Generically, a crossover refers to when a system undergoes a change in phase without encountering a canonical phase transition characterized by a critical point as there are no discontinuities in derivatives of the free energy (as determined by Ehrenfest classification) or symmetry-breaking mechanisms (as determined by Landau classification). A well known example is the BEC-BCS crossover in an ultracold Fermi gas in which tuning the interaction strength (the s-wave scattering length) causes the system to crossover from a Bose-Einstein-condensate state to a Bardeen-Cooper-Schrieffer state.\cite{bec_bcs} Additionally, the Kondo Effect is important in certain metallic compounds with dilute concentrations of magnetic impurities that cross over from a weakly-coupled Fermi liquid phase to a local Fermi liquid phase upon reducing the temperature below some threshold.\cite{kondo} Furthermore, examples of strong crossover phenomena have also been recently discovered in classical models of statistical mechanics such as the Blume-Capel model and the random-field Ising model.\cite{crossover_bcm,crossover_rfim}

The organization of this work is as follows. The next section details the data science and ML methods explored in this work. In section 3, the results of the analysis of the 2-dimensional square Ising model are reported. Section 4 concludes this work with a discussion of the interpretation, implications, and greater impacts of these findings.

\section{Methods}

The Ising configurations are generated using a standard Monte Carlo algorithm written in Python using the NumPy library.\cite{python, numpy} The algorithm was also optimized to be parallel using the Dask library and select subroutines were compiled at run-time for efficiency using the JIT compiler provided by the Numba library.\cite{dask, numba} The Monte Carlo moves used are called spin-flips. A single spin flip attempt consists of flipping the spin of a single lattice site, calculating the resulting change in energy $\Delta E$, and then using that change in energy to define the Metropolis criterion $\exp(-\frac{\Delta E}{T})$. If a randomly generated number is smaller than said Metropolis criterion, the configuration resulting from the spin-flip is accepted as the new configuration. The data analyzed in this work consists of 1,024 square Ising configurations of side length 32 with periodic boundary conditions across 65 external field strengths and 65 temperatures respectively uniformly taken from $\qty[-2, 2]$ and $\qty[1, 5]$. The interaction energies were set to unity such that $J_{ij} = J = 1$. Each sample was equilibrated with 8,192 Monte Carlo updates before data collection began. Data was then collected at an interval of 8 Monte Carlo updates for each sample up to a sample count of 1,024. At the end of each data collection step, a replica exchange Markov chain Monte Carlo move was performed across the full temperature range for each set of Ising configurations that shared the same external field strength.\cite{remcmc,parallel_tempering1,parallel_tempering2} This allows for more robust sampling of the ensemble across the temperature range by allowing high-temperature states to be available at low temperatures as well as the inverse. Additionally, this helps to prevent samples on the vanishing field line from relaxing into either positive or negative magnetization states since two states at temperatures close to one another opposite spins would be very likely to swap.

In this work, the  Ising spins were rescaled such that a spin-down atomic spins carry the value 0 and spin-up atomic spins carry the value 1, which is a standard setup for binary-valued features in data science. Physically, this would be interpreted as the lattice gas model as described in the prior section.

The goal is to map the raw Ising configurations to a small set of descriptors that can discriminate between the samples using a structural criterion inferred by an ML algorithm. This application is referred to as representation learning and is often presented as dimensionality reduction. There are many methods in the field of unsupervised ML that seek to achieve such data dimensionality reduction\cite{pca,tsne} however, such methods do not respect the multidimensional structure of the input data, so a deep neural network will be used instead to accomplish the data dimensionality reduction in the form of a self-supervised variational autoencoder (VAE).\cite{vae} Such a neural network is composed of three main components, an encoder network, a decoder network, and a sampling function. The encoder and decoder neural networks are implemented as deep convolutional neural networks (CNN) in order to preserve the spatially dependent 2-dimensional structure of the Ising configurations.\cite{cnn} The general idea of a VAE is to encode configurations into a latent variable space composed of the parameters for a chosen prior distribution. A multidimensional Gaussian distribution was used for this work. Random variables from these distributions can then be decoded to recover the original input configurations. In this way, VAEs are both generative models and latent variable models. Assuming a model is sufficiently trained, new sample data can be generated through traversing the latent space input to the decoder network.

The purpose for using a VAE in this manner is to extract a low-dimensional representation of the Ising configurations that are otherwise unwieldy to compare directly in a meaningful manner without \textit{a priori} knowledge of the important derived measurements from statistical physics used to accomplish the same tasks. The motivation then for using a VAE to encode and decode the Ising configurations lies in the desire to automate the parameterization of the Ising configurations without conventional methods from statistical physics, preferring instead to allow the neural network to learn and discover the important features itself directly from the structures of the configurations. The latent representations of the configurations will be small sets of descriptors for the configurations that can be used to discriminate between them by relying on the assumption that proximities between latent representations of the Ising configurations in the latent space are notions of structural similarity between the configurations in their original 2-dimensional lattice representations. In this way, the VAE is is used as an alternative to conventional statistical mechanics algorithms to accomplish the same task of characterizing the structural features of input configurations.

The encoder CNN uses four convolutional layers with kernel shapes of $\qty(3, 3)$ following the input layer with kernel strides of $\qty(2, 2)$ and increasing filter counts by a factor of 4. Zero-padding is used to ensure that the entire input is reached with convolutions. Furthermore, each convolutional layer uses scaled exponential linear unit activation functions (SELU) and LeCun normal initializations as well as kernels of shape $\qty(3, 3)$.\cite{snn} The output of the final convolutional layer is then flattened, feeding into two dense layers of eight neurons respectively representing the latent variables that correspond to the means $\mu_i$ and logarithmic variances $\log\sigma_i^2$ of multivariate Gaussian distributions using linear activations. A random variable $z_i$ is drawn from the distribution such that $z_i = \mu_i+\exp[\frac{1}{2}\log\sigma_i^2]\mathrm{N}_{0,1}$, where $\mathrm{N}_{0,1}$ is the standard normal distribution. The logarithmic variance is used in favor of the standard deviation directly in the interest of maintaining numerical stability. The random variable $z_i$ is then used as the input layer for the decoder CNN, where $z_i$ is mapped to a dense layer that is then reshaped to match the structure of the output from the final convolutional layer in the encoder CNN. From there, the decoder CNN is simply the reverse of the encoder network in structure, albeit with convolutional transpose layers in favor of standard convolutional layers. The final output layer from the decoder network is thus a reproduction of the original input configurations to the encoder network using a sigmoid activation function. The structures of the VAE as a whole is shown in Fig. 1 as well as an example of a convolution operation that composes the bulk of the operations in the encoder and decoder networks is shown in Fig. 2.

\begin{figure}[H]
\centering
\includegraphics[width=0.8\linewidth]{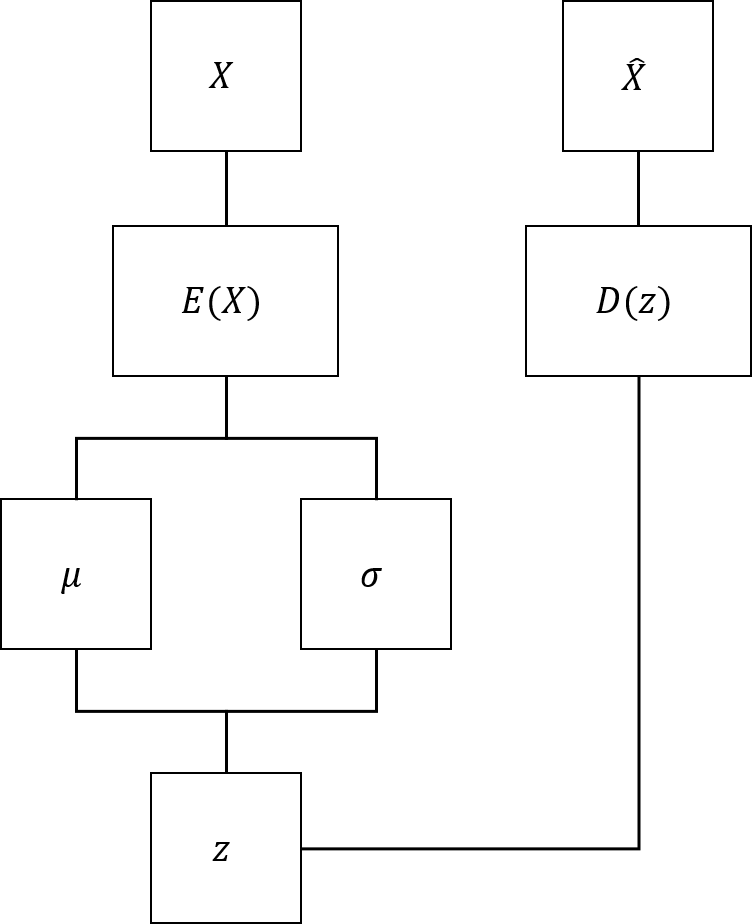}
\caption{A diagram depicting the structure of the VAE where $X$ is the input Ising configuration, $E(X)$ is the encoder network, $\mu$ and $\sigma$ are the latent means and standard deviations, $z$ is the random Gaussian sample from the distribution described by $\mu$ and $\sigma$, $D(z)$ is the decoder network, and $\hat{X}$ is the reconstructed Ising configuration.}
\end{figure}

\begin{figure}[H]
\centering
\includegraphics[width=0.8\linewidth]{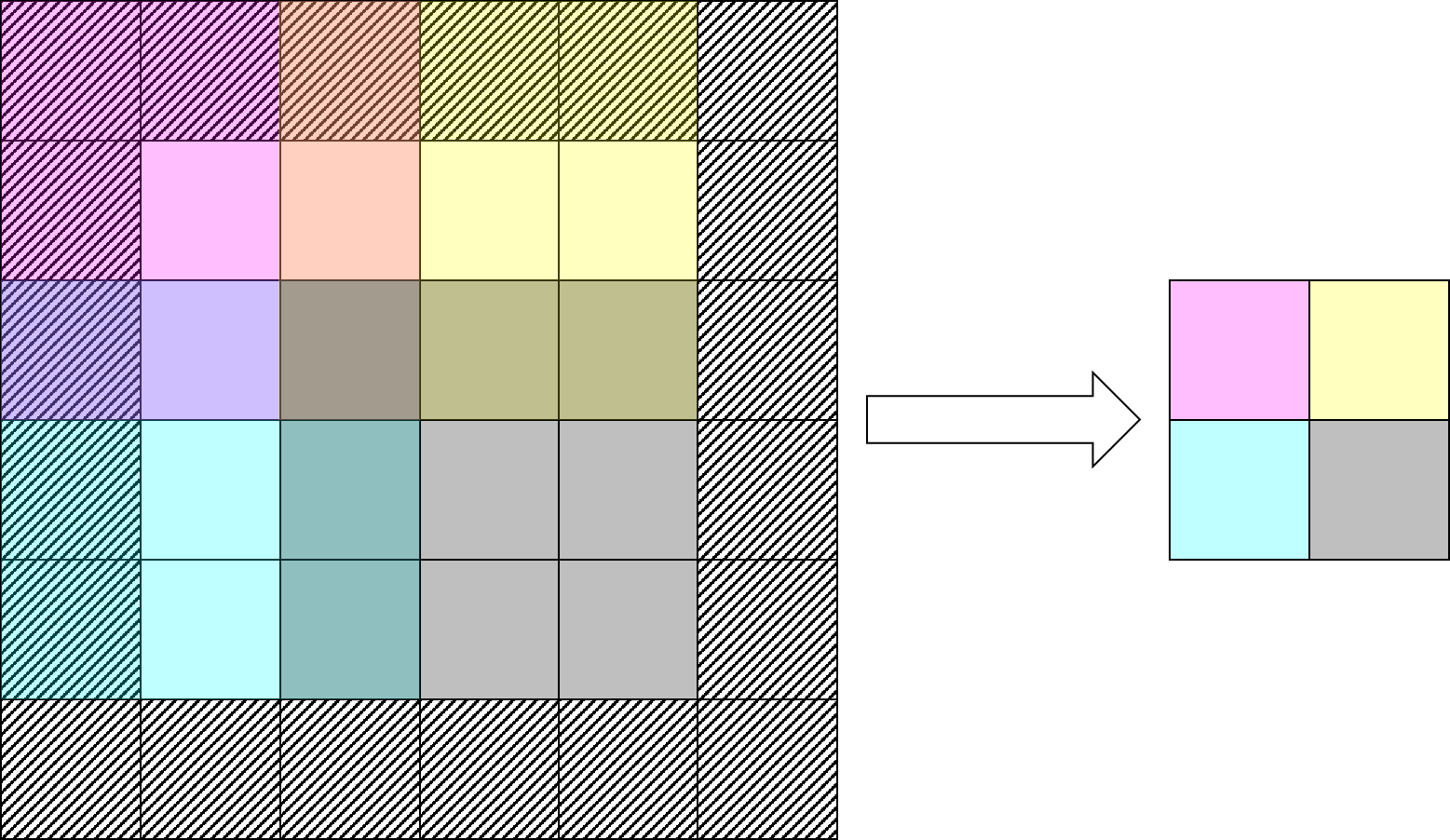}
\caption{A diagram depicting the convolution operation for a single kernel of shape (3, 3) with a stride of (2, 2) acting on an input of shape (4, 4) with zero-padding denoted by the striped input region to produce an output feature map of shape (2, 2). Each stride is color coded such that each entry in the output is the sum of the products of the kernel weights and input entries over the subvolume corresponding to the same color. Since the stride is less than the kernel size, the subvolumes overlap.}
\end{figure}

The loss term consists of two separate components. The first is the standard reconstruction loss, which was implemented using the binary crossentropy between the encoder input and decoder output in this work. Other choices for the reconstruction loss are still valid, however, such as mean squared error or mean absolute error. The second loss term is a Kullback-Liebler divergence term which acts as a regularizer to ensure the latent variables $\mu_i$ an $\sigma_i$ faithfully represent multivariate Gaussian parameters. The combination of the reconstruction loss and the Kullback-Leibler divergence is called the tractable evidence lower bound, often referred to as ELBO. In this work, the Kullback-Leibler term was decomposed in a manner similar to a $\beta-$ total correlation VAE ($\beta-$TCVAE) network, separating it into three parts describing the index-code mutual information, total correlation, and dimension-wise Kulback-Leibler divergence.\cite{btcvae} Minibatch stratified sampling was also employed during training.\cite{btcvae} The specific parameters of the decomposition used were $\alpha=\lambda=1$ and $\beta=8$.

The Nesterov-accelerated Adaptive Moment Estimation (Nadam) optimizer was used to optimize the loss, though many other choices are available.\cite{nadam} It was found that the adaptive nature of the Nadam optimizer more efficiently arrived at minimizing the loss during training of the $\beta-$TCVAE model than other optimizers. The specific parameters used for the Nadam optimizer were $\beta_1 = 0.9$, $\beta_2 = 0.999$, a schedule decay of $0.4$, and the default epsilon provided by the Keras library. A learning rate of $0.00001$ was chosen. Training was performed over $16$ epochs with a batch size of $845$ and the samples were shuffled before training started. A callback was used to reduce the learning rate on a loss plateau with a patience of $8$ epochs.

After fitting the $\beta-$TCVAE model, the latent encodings of the Ising configurations were extracted for further analysis. Principal component analysis (PCA) was used on the latent means and standard deviations independently to produce linear transformations of the Gaussian parameters that more clearly discriminate between the samples using the scikit-learn package in an attempt to further disentangle the representations provided by the $\beta$-TCVAE.\cite{pca} This is done by diagonalizing the covariance matrix of the original features to find a set of independent orthogonal projections that describe the most statistically varied linear combinations of the original feature space.\cite{pca} The PCA projections are then interpreted for the 2-dimensional Ising model. The motivation for using the principal components (PC) of the latent variables instead of the raw latent variables is to more effectively capture measurements that are both statistically independent due to the orthogonality constraint and also explain the most variance possible in the latent space under said constraint. Given that the latent representations characterize the structure of the Ising configurations, the principal components of the latent representations allow for more effective discrimination between the different structural characteristics of the configurations than the raw latent variables do. The $\beta-$TCVAE model used in this work was implemented using the Keras ML library with TensorFlow as a backend.\cite{keras, tf}

\section*{Results}

All of the plots in this section were generated with the MatPlotLib package using a perceptually uniform colormap.\cite{matplotlib} In each plot, the coloration for a square sector on the diagram represents the average value of the measurement at that sector on the diagram, with lightness corresponding to magnitude.

The latent means $\mu_i$ contain only one PC with noticeable statistical significance as it explains 77.1\% of the total statistical variances between the $\mu_i$ encodings of the Ising configurations while the rest explained less than 4\% each. This PC will be denoted with $\nu_0$. These results reflect the accomplishments of prior published works.\cite{carrasquilla2017machine}

\begin{figure}[H]
\centering
\includegraphics[width=0.8\linewidth]{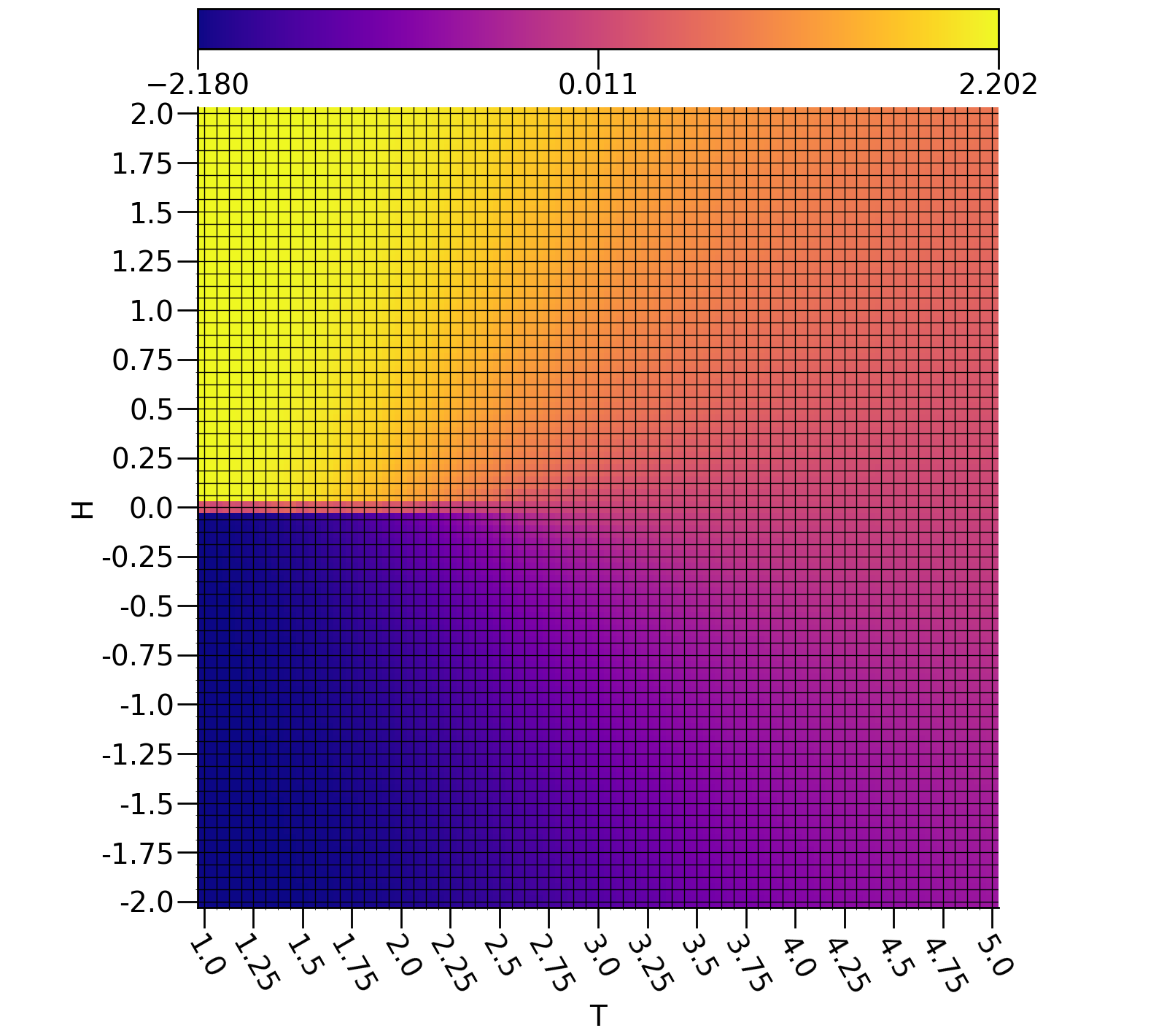}
\caption{The ensemble average $\nu_0$ with respect to the external field strengths and temperatures.}
\end{figure}

\begin{figure}[H]
\centering
\includegraphics[width=0.8\linewidth]{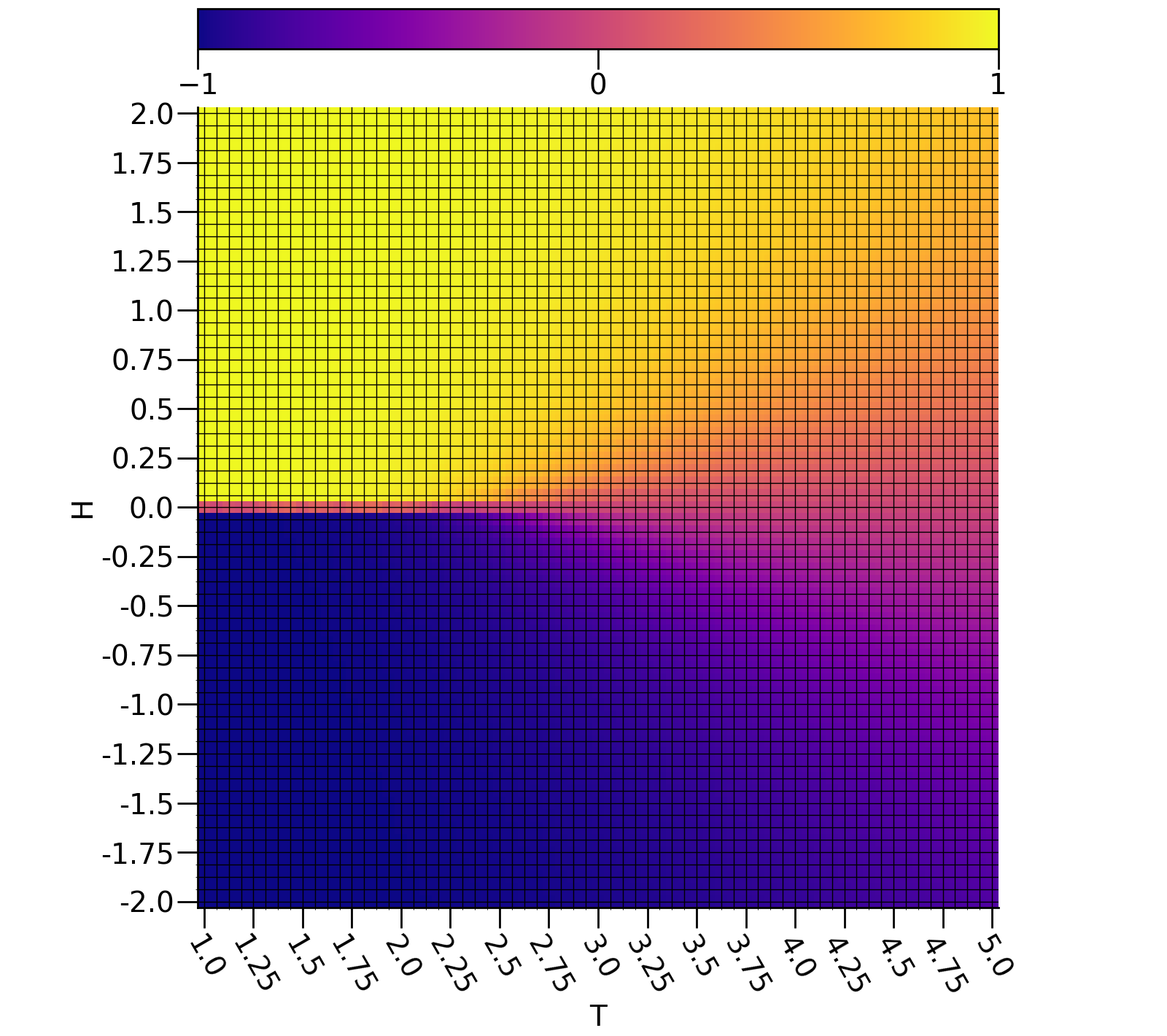}
\caption{The ensemble average magnetization $m$ with respect to the external field strengths and temperatures.}
\end{figure}

By comparing $\nu_0$ depicted in Fig. 3 to the calculated magnetizations $m$ of the Ising configurations in Fig. 4, it is readily apparent that $\nu_0$ is rather faithfully representing the magnetizations of the Ising configurations. There are some inaccuracies in the intermediate magnetizations produced by a relationship resembing a sigmoid between $\nu_0$ and $m$, but a very clear discrimination between the ferromagnetic spin-up and ferromagnetic spin-down configurations is shown. Since the magnetizations act as the order parameter for the 2-dimensional Ising model, this shows that the extraction of a reasonable representation of the order parameter is possible with a VAE. It is important to note that since the magnetization is a linear feature of the Ising configurations, a much simpler linear model would be sufficient for extracting the magnetization.

The latent standard deviations $\sigma_i$ show much more interesting behavior, however. Two PCs of the $\sigma_i$ encodings, denoted as $\tau_0$ and $\tau_1$, are investigated with respect to the external field strengths and the temperatures.

\begin{figure}[H]
\centering
\includegraphics[width=0.8\linewidth]{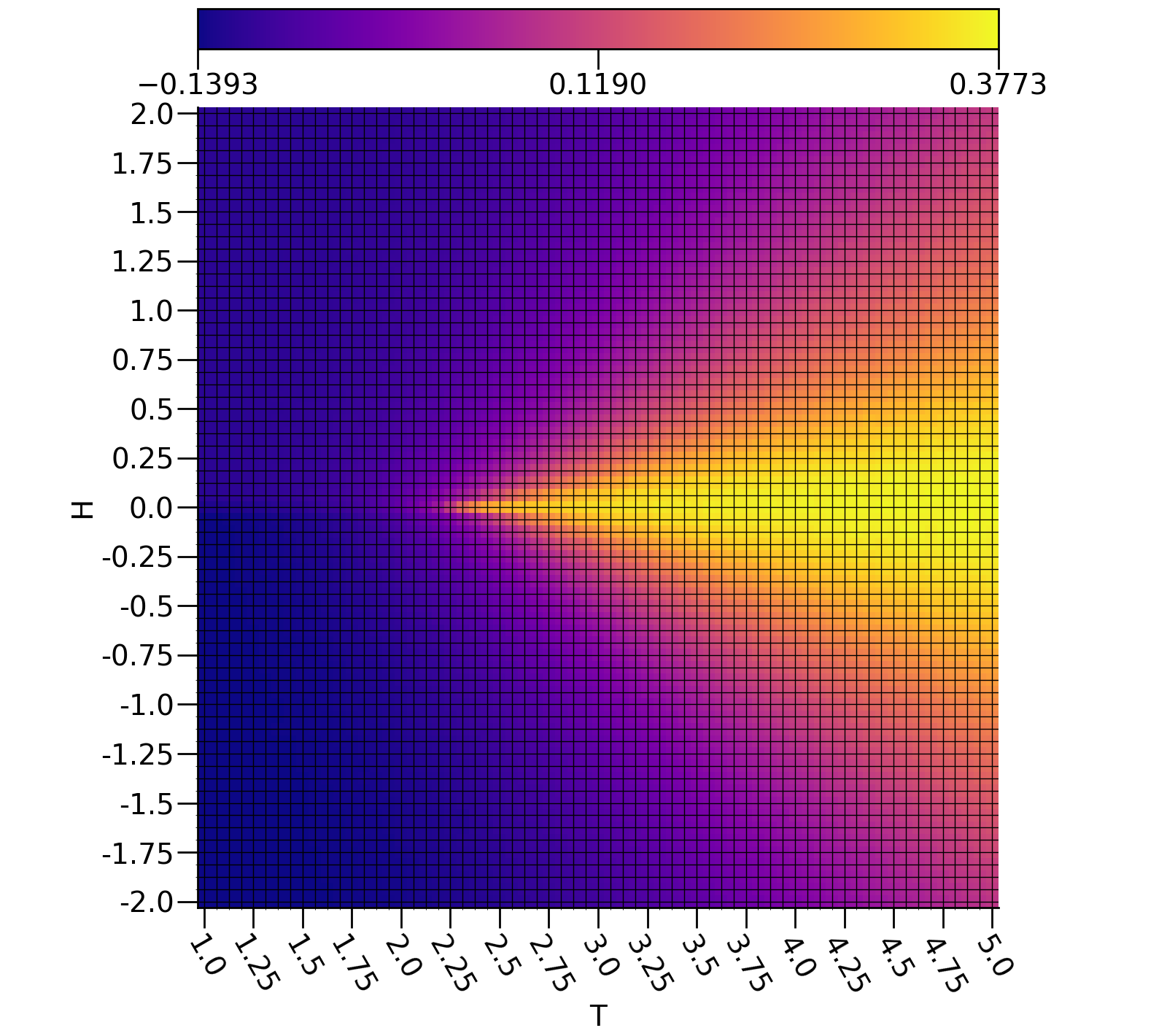}
\caption{The ensemble average $\tau_0$ with respect to the external field strengths and temperatures.}
\end{figure}

\begin{figure}[H]
\centering
\includegraphics[width=0.8\linewidth]{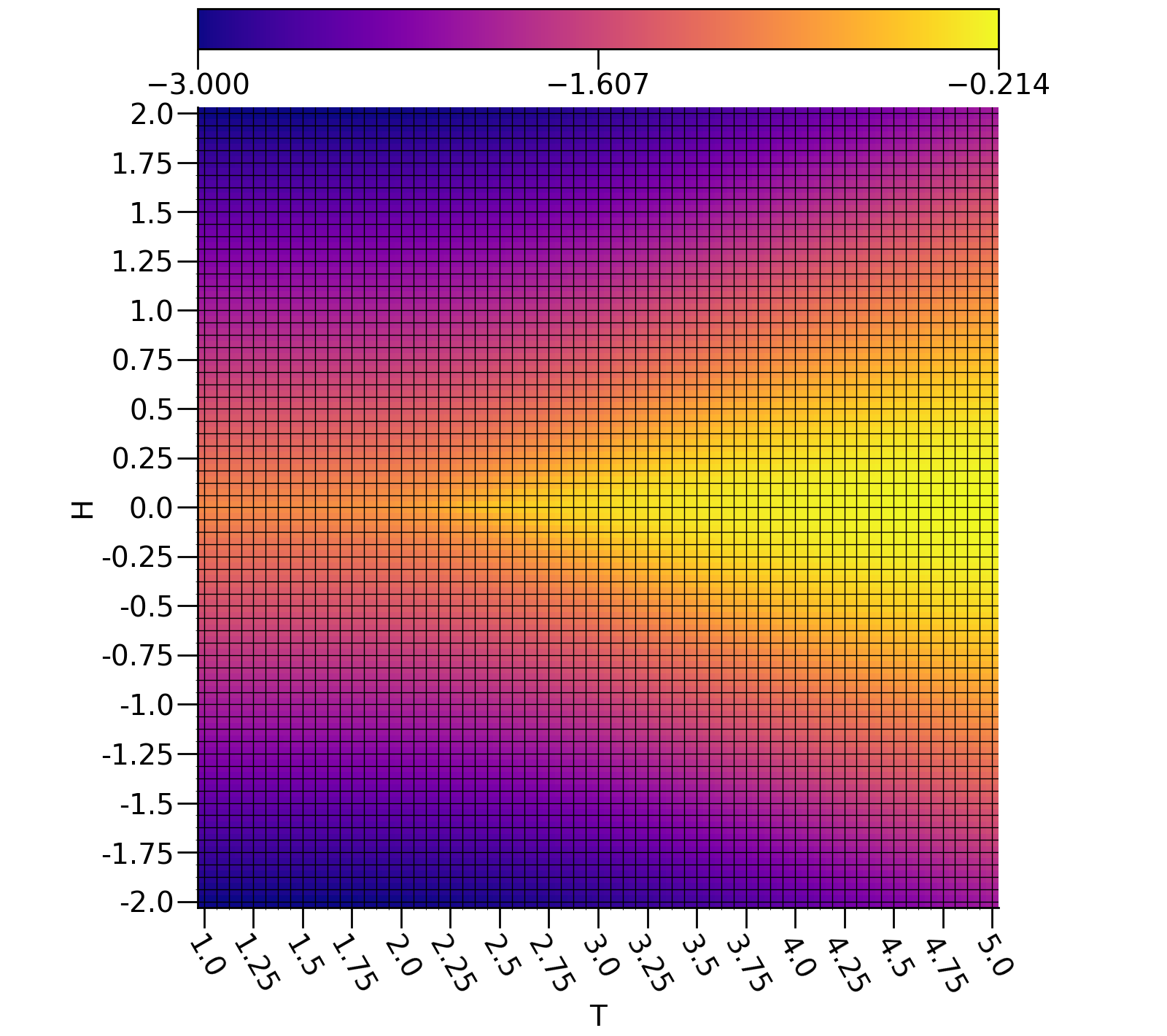}
\caption{The ensemble average energy $E$ with respect to the external field strengths and temperatures.}
\end{figure}

By comparing $\tau_0$ depicted in Fig. 5 to the calculated energies $E$ of the Ising configurations shown in Fig. 6, it is clear that $\tau_0$ exhibits a strong discrimination between the low to intermediate energy regions and the highest energy region characterized by a cone starting at the vanishing field critical point approximated at $T_C \approx 2.25$ that extends symmetrically to include more external field values with rising temperature, which is rather similar to the critical point predicted using a dense autoencoder.\cite{ising_vae_trans} This is in effect capturing the concretely paramagnetic samples and the relative error in the estimation of the critical temperature is acceptable with a 0.85\% overestimate error with respect to the exact value of $T_C = \frac{2}{\ln\qty[1+\sqrt{2}]} \approx 2.27$.\cite{PhysRev.65.117} Given that the paramagnetic samples are essentially noise due to entropic contributions from thermal fluctuations destroying any order that would otherwise be present, it makes sense that these would be easy to discriminate from the rest of the samples using a $\beta-$TCVAE model. This is because the samples with $\nu_0$ values corresponding to nearly zero magnetizations and rather high values for $\tau_0$ will resemble Gaussian noise with no notable order preference, which is indeed reflected in the raw data. In this way, it seems that $\nu_0$ is suitable for tracking the ferromagnetic ordering while $\tau_0$ is suitable for characterizing the paramagnetic disorder.

\begin{figure}[H]
\centering
\includegraphics[width=0.8\linewidth]{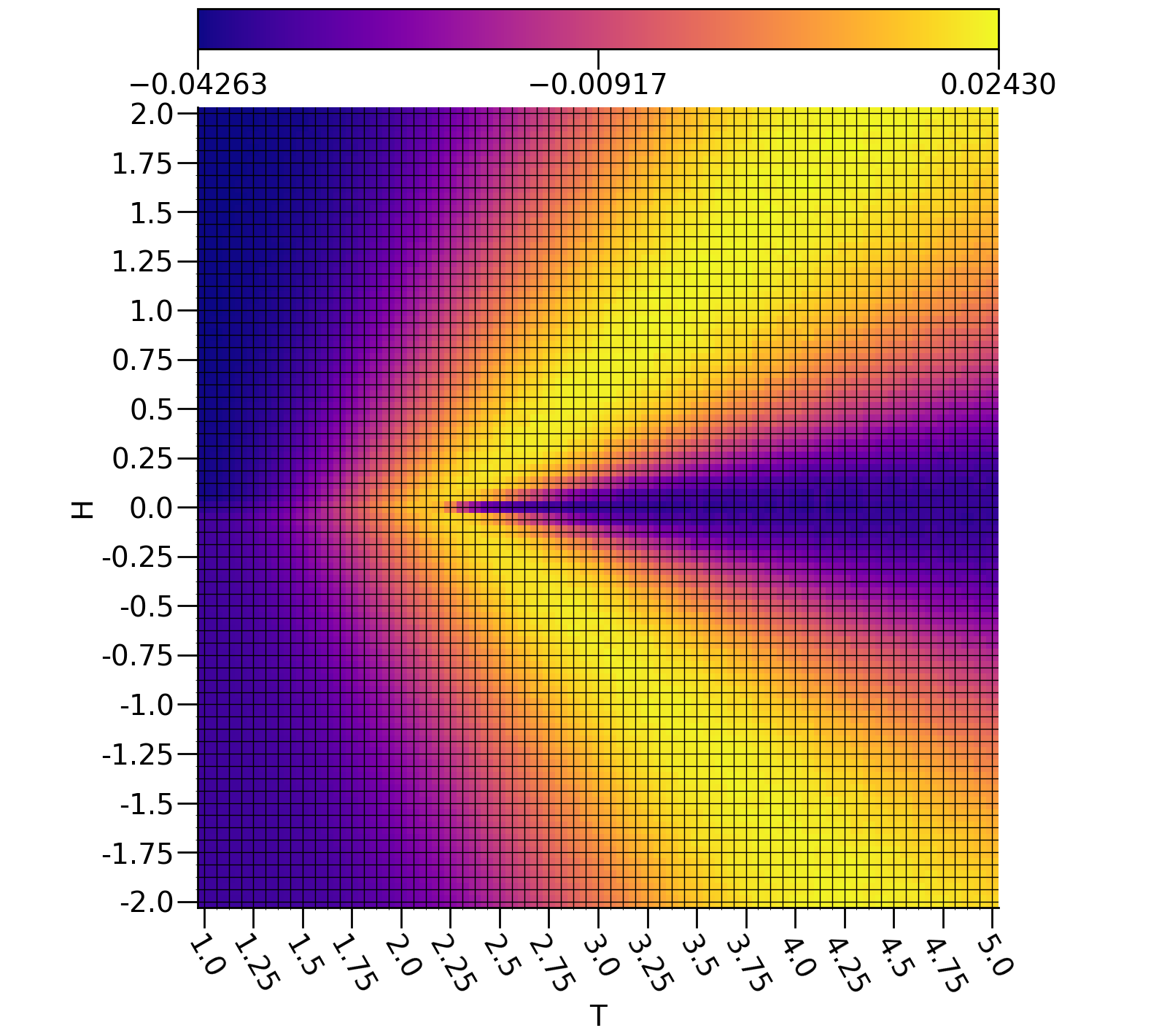}
\caption{The ensemble average $\tau_1$ with respect to the external field strengths and temperatures.}
\end{figure}

\begin{figure}[H]
\centering
\includegraphics[width=0.8\linewidth]{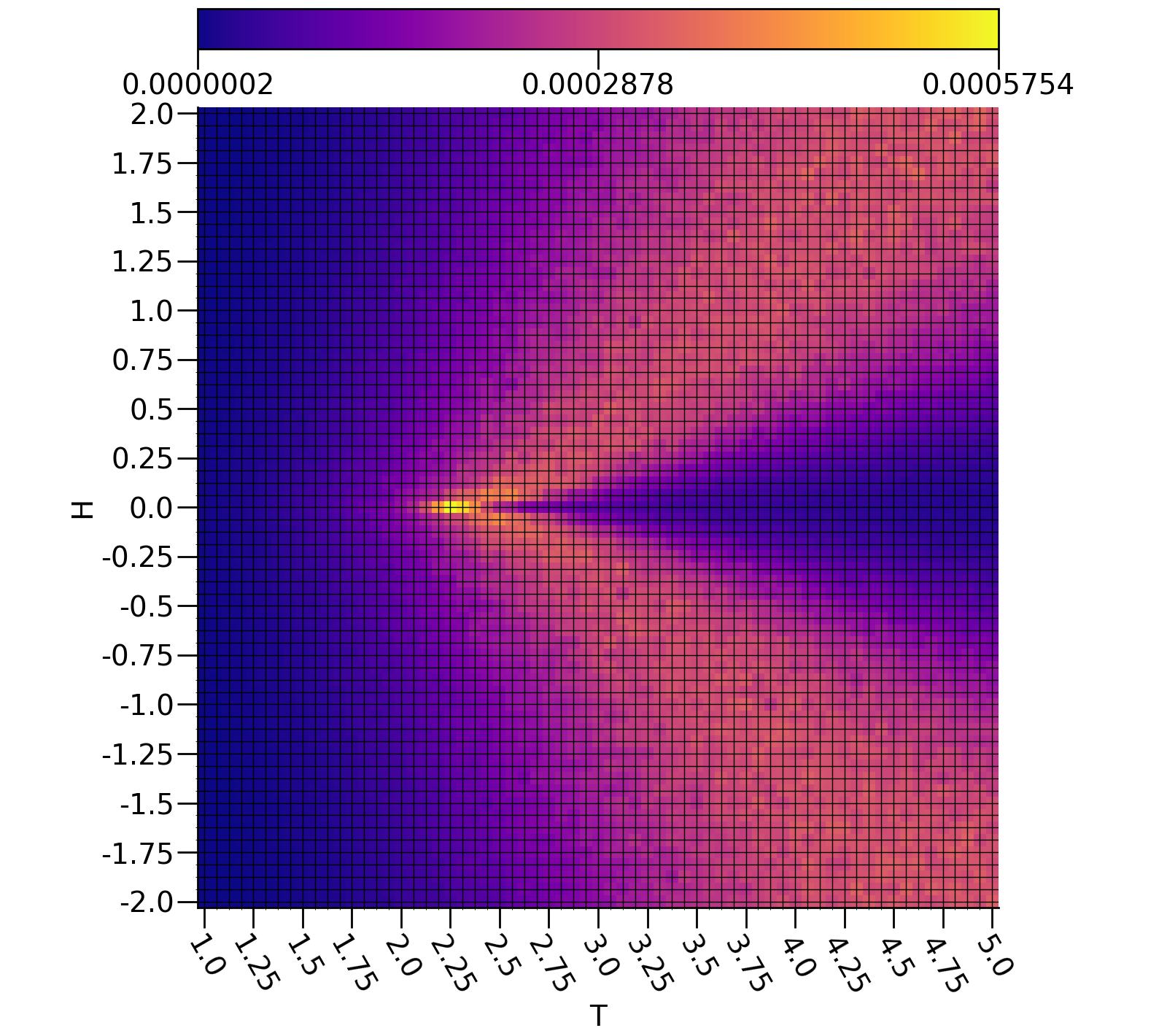}
\caption{The ensemble Ising specific heat $C$ with respect to the external field strengths and temperatures.}
\end{figure}

The behavior of $\tau_1$ shown in Fig. 7 is even more interesting, as it is not simply discriminating samples with intermediate energies from the rest of the data set. If this were true, then some samples at temperatures below the critical point at non-zero external field strengths would be included, as is readily apparent in the energies shown in Fig. 6. Rather, there is another cone shape as was seen with $\tau_0$, albeit much wider and with the the samples represented strongly by $\tau_0$ omitted. In effect, it would appear as if $\tau_1$ is capturing regions in the diagram with intermediate structural disorder as opposed to the highly disordered structures captured by $\tau_0$. Interestingly, $\tau_1$ bears a rather strong resemblance to the specific heat capacity $C$ depicted in Fig. 8. It is worth noting that there is a slight asymmetry between the spin up and the spin down configurations in $\tau_1$, but it has negligible effects on the relevant analysis.

\begin{figure}[H]
\centering
\includegraphics[width=0.8\linewidth]{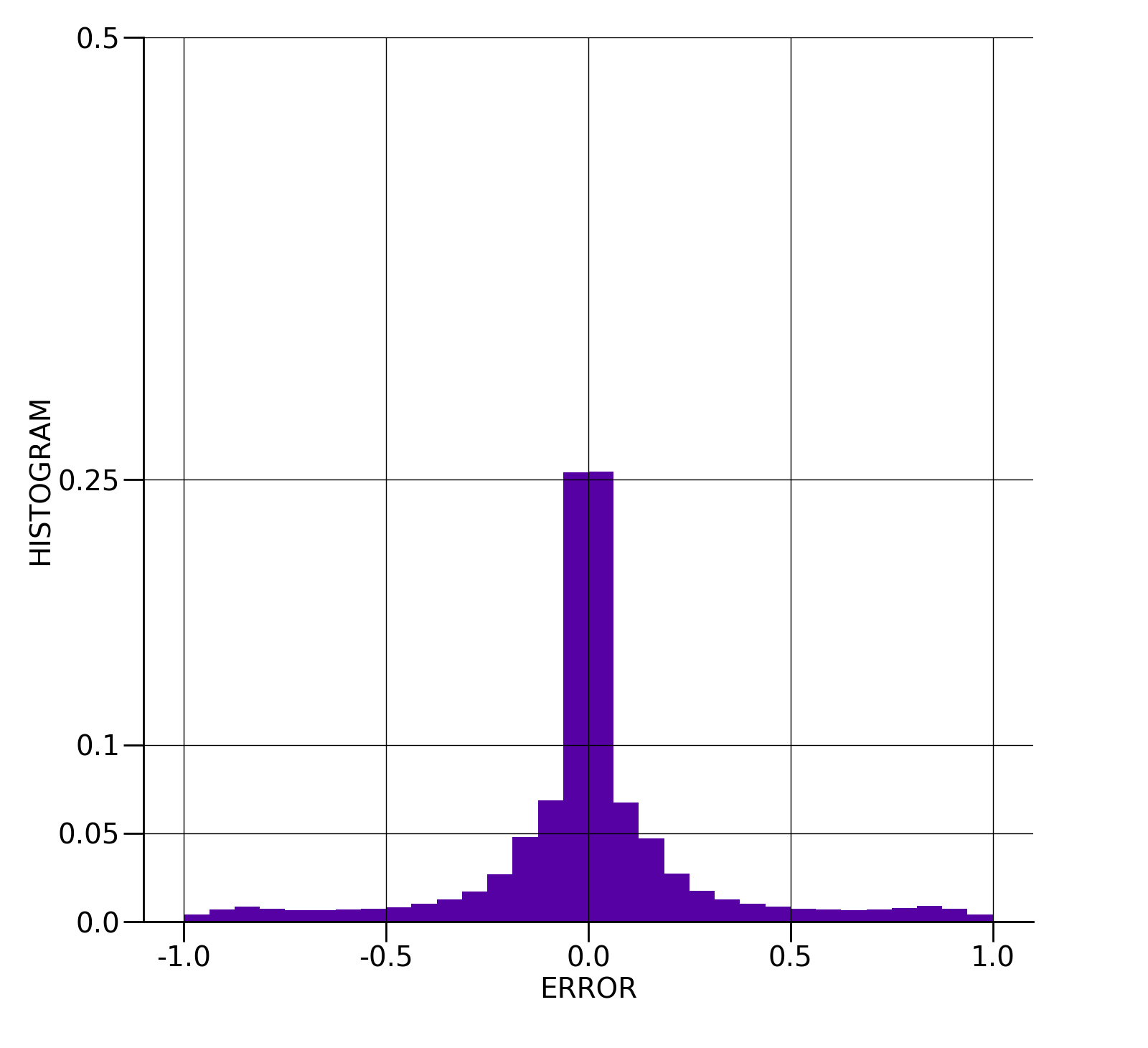}
\caption{The distribution of errors in the $\beta-$TCVAE Ising spin predictions.}
\end{figure}

\begin{figure}[H]
\centering
\includegraphics[width=0.8\linewidth]{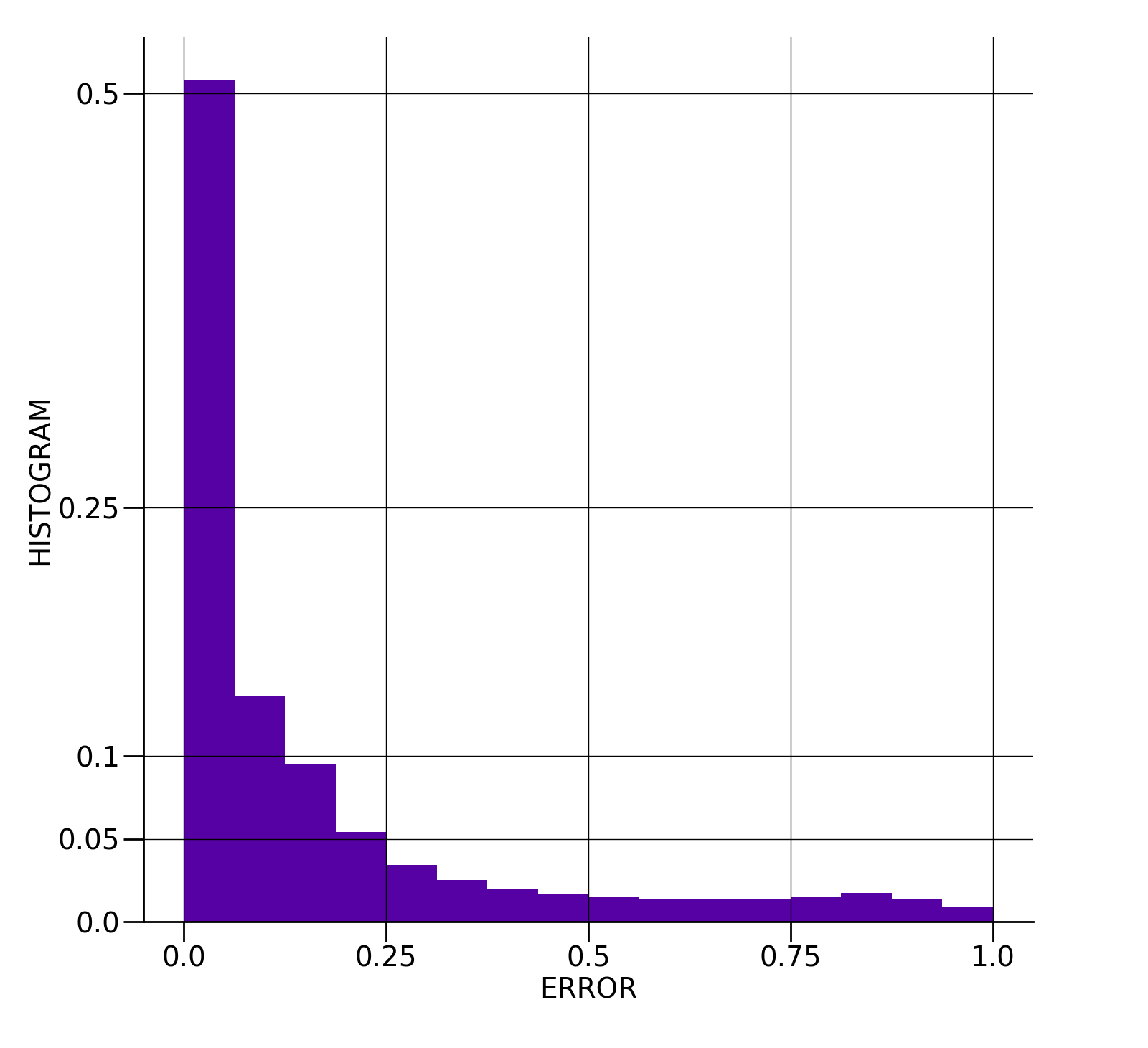}
\caption{The distribution of absolute errors in the $\beta-$TCVAE Ising spin predictions.}
\end{figure}

\begin{figure}[H]
\centering
\includegraphics[width=0.8\linewidth]{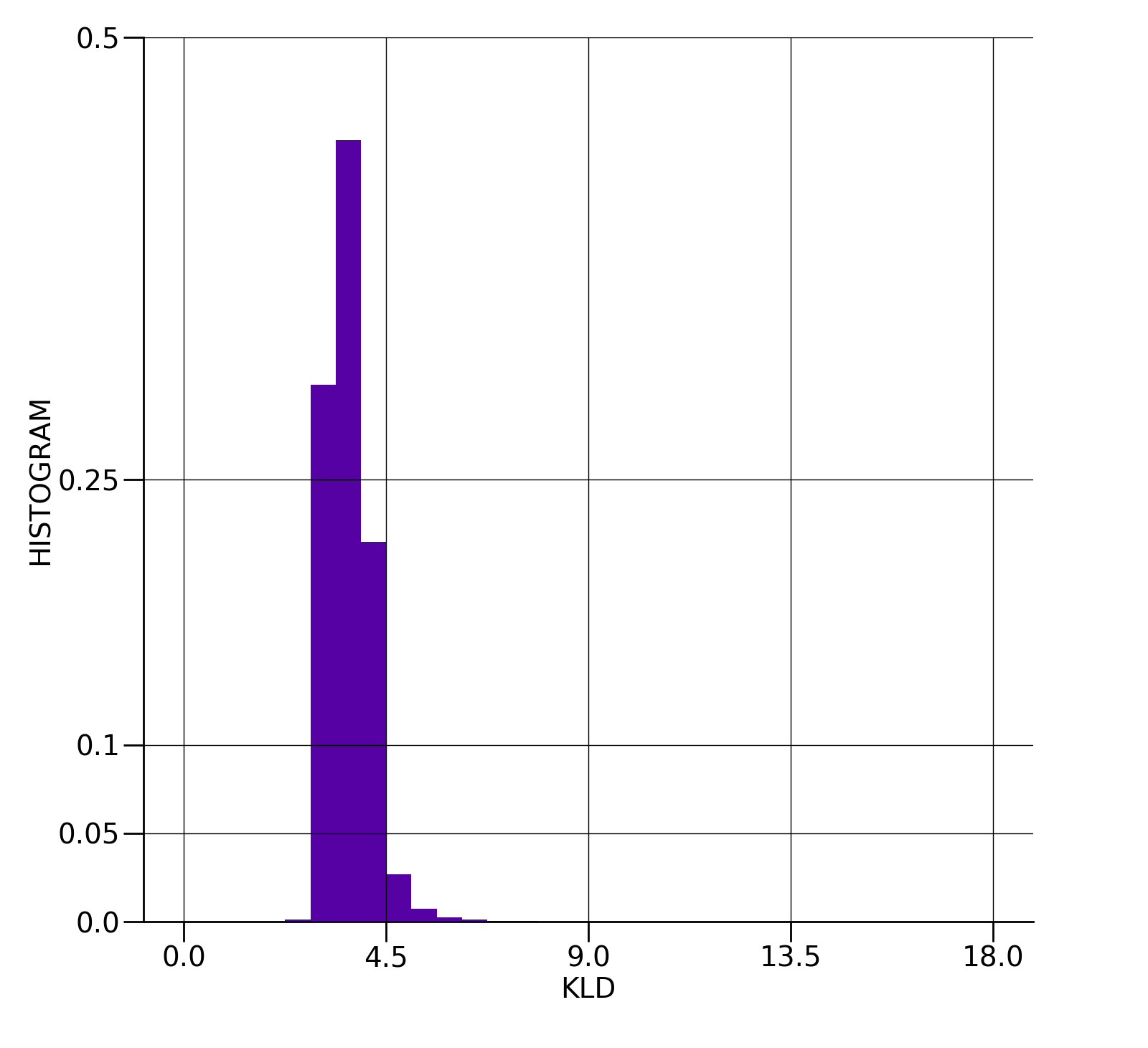}
\caption{The distribution of Kullback-Leibler divergences of the $\beta-$TCVAE model latent encodings.}
\end{figure}

The distribution of the error between the true and $\beta-$TCVAE predicted values for the Ising model spins is shown in Fig. 9. The distribution is very sharply centered around and reasonably symmetric about zero, showing suitable spin prediction accuracy without a considerable bias towards one spin over the other. The distribution of the absolute errors between the true and $\beta-$TCVAE predicted values is shown in Fig. 10, showing that the bulk of the predictions exhibit very little error. The distribution of the Kullback-Leibler divergences is depicted in Fig. 11 and is well-behaved with few outliers.

\section{Discussion}

In essence, using a VAE to extract structural information from raw Ising configurations exposes interesting derived descriptors of the configurations that can be used to not only identify a transition point, but also a crossover region amongst other regions of interest. The crux of this analysis is in the interpretation of the extracted feature space as represented by the latent variables. This is done by studying the behavior of the latent variable mappings of the Ising configurations with respect to the external magnetic fields and temperatures.

Considering that $\nu_0$ reflects the magnetization for the 2-dimensional Ising model, this means it can be readily interpreted as an indicator for the ferromagnetic ordering exhibited by the configurations. By contrast, $\tau_0$ and $\tau_1$ can be interpreted as an indicator of paramagnetic disorder that also provides a suitable estimate of the transition temperature. The extracted region from $\tau_1$ can readily be interpreted as the crossover region, as these configurations exhibit order preferences alongside a significant amount of noise brought on by the entropic contributions from the thermal fluctuations at higher temperatures. As would be expected of the crossover region, it shifts to higher temperatures with increasing external magnetic field strengths.

These results potentially carry broad implications for the path towards formulating a generalized order parameter alongside a notion of a crossover region with minimal \textit{a priori} information through the use of ML methods, which would allow for the investigation of many interesting complex systems in condensed matter physics and materials science. The advantage of the present method is in its capability of capturing the crossovers. This opens a new avenue for the study of quantum critical points from the data obtained at low but finite temperatures that instead exhibits crossover regions. Examples of these include data from large scale numerical Quantum Monte Carlo simulations for heavy fermion materials and high temperature superconducting cuprates for which quantum critical points are believed to play crucial roles for their interesting properties.\cite{qcp,singular_fl,2D_DCA_QCP}

There are many opportunities beyond investigating more complex systems by introducing improvements to this method beyond the scope of this work. For instance, finite-size scaling is an important approach towards addressing limitations presented by finite-sized systems for investigation critical phenomena.\cite{fsc} Establishing correspondence between the VAE encodings of different system sizes is a challenging proposition, as different VAE structures will need to be trained for each system size, which in turn may require different hyperparameters and training iteration counts to provide similar results. Consequently, numerical difficulties can arise when performing finite-size scaling analysis, as the variation of predicted properties with respect to system size may be difficult to isolate from the systemic variation due to different neural networks being used to extract said properties. Nevertheless, this would be a significant step towards improving VAE characterization of critical phenomena.

\section{Acknowledgements}

This work is funded by the NSF EPSCoR CIMM project under award OIA-1541079. Additional support (MJ) was provided by NSF Materials Theory grant DMR-1728457. An award of computer time was provided by the INCITE program. This research also used resources of the Oak Ridge Leadership Computing Facility, which is a DOE Office of Science User Facility supported under Contract DE-AC05-00OR22725.

\bibliography{refs}
\bibliographystyle{apsrev}

\end{document}